\documentclass[useAMS,usenatbib]{mn2e}

 \usepackage{times}
 \usepackage{graphicx}
 \usepackage{journals}
 \usepackage{soul}

\newcommand{\km}{\,\mbox{km}\,\mbox{s}$^{-1}$} 

\def\Ha{H$\alpha$ \,} 
\def\Hb{H$\beta$ \,} 
\def\i{\,{\small I}}
\def\ii{\,{\small II}} 
\def\iii{\,{\small III}}

\title[Decoupled gas kinematics]{Decoupled gas kinematics in isolated S0 galaxies}
\author[Katkov, Sil'chenko, Afanasiev]
{Ivan Yu. Katkov$^{1,2}$\thanks{E-mail: katkov.ivan@gmail.com}, Olga K. Sil'chenko$^{1,3}$,
\and Victor L. Afanasiev$^{4}$\\
$^{1}$Sternberg Astronomical Institute, Lomonosov Moscow State University, Universitetski prospect 13, 119992 Moscow, Russia\\
$^{2}$Faculty of Physics, Lomonosov Moscow State University, Leninskie gory 1, 119991, Moscow, Russia\\
$^{3}$Isaac Newton Institute of Chile, Moscow Branch, Universitetski prospect 13, 119992, Moscow, Russia\\
$^{4}$Special Astrophysical Observatory, Nizhnij Arkhyz, 369167,  Russia 
}

\begin{document}

\date{Accepted 2013 ????? ??. Received 2013 ????? ??; in original form 2013 December 24}

\pagerange{\pageref{firstpage}--\pageref{lastpage}} \pubyear{2013}

\maketitle

\label{firstpage}

\begin{abstract} 
A sample of completely isolated S0 galaxies has been studied by means of
long-slit spectroscopy at the Russian 6-m telescope. 7 of 12 galaxies
have revealed a presence of extended ionized-gas discs which rotation is mostly
decoupled from the stellar kinematics: 5 of 7 (71$\pm$17\%) galaxies show a
visible counterrotation of the ionized-gas component with respect to the stellar
component. The emission-line diagnostics demonstrates a wide range of the gas
excitation mechanisms, although a pure excitation by young stars is rare. We
conclude that in all cases the extended gaseous discs in our sample S0s are of
external origin, despite the visible isolation of the galaxies.  Possible
sources of external accretion, such as systems of dwarf gas-rich satellites or
cosmological cold-gas filaments, are discussed.
\end{abstract}


\section{Introduction}

One of the central topics in current extragalactic astronomy is how galaxies
form and how their properties change through cosmic times. This is a
particularly difficult task if concerning lenticular galaxies since this class
of objects show a diversity in their properties that goes beyond present day
simulations.  A lot of internal and external physical processes controlling
evolution of galaxies through cosmic time may, and must, play role.  External
processes -- gravitational tides, major and minor mergers, dry and wet ones,
external cold gas accretion through intergalactic medium, ram pressure in hot
intracluster/intragroup medium -- govern star formation and re-shape general
structures.  Internal disc instabilities may provoke secular evolution and
matter radial re-distribution.  The main point is to study galaxies where
manifold of processes is constrained and so evolution is only driven by few main
processes. Such are isolated galaxies.
 
We have compiled our sample of isolated S0 galaxies basing on the approaches
recently developed by the team of Karachentsev, Makarov, and co-authors.  Their
group-finding algorithm that takes into account individual characteristics of
galaxies has been already used to study the properties of isolated galaxies
\citep{Karachentsev2011LOG}, binary \citep{Karachentsev2008binary}, triple
\citep{Makarov2009triplets} systems of galaxies, and galaxy groups
\citep{Makarov2011groups}.  Our sample objects, 18 S0 galaxies which are rather
nearby, $V_r<3500$ km/s, and relatively luminous, $M_K<-19$, satisfy the
following criterion of isolation: isolation index $\kappa>2.5$. We have
undertaken long-slit spectroscopy of 12 targets at the Russian 6m telescope; 7
of them have revealed extended emission lines; and among those, 5 galaxies
demonstrate decoupled gas kinematics with respect to their stellar components.
In the remaining five galaxies the emission lines are undetectable.  Previously,
\citet{atlas3d_10} studied three galaxies of our list using their IFU data; they
mentioned kinematical misalignment of stars and ionized gas in these galaxies. 

Although lenticular galaxies typically have less amounts of cold gas than spiral
galaxies, it is now becoming clear that atomic and/or molecular gas is present
perhaps in most of them \citep{welchsage03,
welchsage06,welchsage10,Young2013_atlas3d4_H2}, though less than the half of S0
galaxies with extended cold-gas discs experience current star formation
\citep{pogge_esk93}.  Frequent incidence of decoupled gas kinematics in S0s
allowed to \citet{bertola92} to conclude that at least 40\%\ of emission-line
S0s acquired their gas from external sources.  Basing on a larger S0 sample,
\citet{zeilinger93} found a half of all nearby S0s with extended ionized-gas
emission possessing decoupled gas kinematics.  However here environment may be
important: a recent study by \citet{atlas3d_10} has shown that among the Virgo
cluster S0s the gaseous and stellar components demonstrate always kinematical
alignment while the non-Virgo S0s, those in groups and in field, have decoupled
gas kinematics in 50\%\ of all cases. Our sample -- the strictly isolated S0s --
represents the extreme case of sparse environment, so if the role of environment
is such as \citet{atlas3d_10} indicate, one could expect a large fraction of
decoupled gas kinematics just within our sample.

This paper is organized in the following way. Section 2 describes our
observations and data analysis approaches. In Section 3 we present our results
on the gas and star kinematics in the S0 galaxies where we have detected
emission lines, and in Section 4 we discuss and conclude.

\begin{table*} 
\begin{minipage}{\textwidth}

\caption[ ] {Global parameters of the isolated S0 galaxies with emission lines} 
\begin{tabular}{lccccccc} 
\hline
Galaxy  & NGC~2350     & NGC~3248  & NGC~6654      & NGC~6798  & NGC~7351  & UGC~4551   & UGC~9519  \\ 
Type (NED$^1$)  
	& S0/a & S0        & (R')SB0/a(s)  & S0        & SAB0$^0$? & S0?        & S0?  \\ 
$R_{25}$, kpc (NED$+$RC3$^2$) 
		& 5.6	 & 9.2 	 & 11.3 	& 7.1	  & 1.9 	  & 7.8        & 2.9 \\ 
$R_{25}$, arcsec (NED$+$RC3) 
	& 40.5	 & 75.35	 & 78.9	& 47.5	  & 53.35 	  & 61.25      & 23.8 \\ 
$B_T^0$ (LEDA$^3$) & 13.19 & 13.80	  & 12.52 	& 13.72 	  & 13.39 	  & 13.27 	   & 14.22 \\ 
$M_K$ (NED+LEDA)$^4$	& --22.73 & --21.82 	 & --23.83	 & --23.52 	  & --20.92        & --22.63   & --21.71 \\ 
$V_r$, $\mbox{km} \cdot 
\mbox{s}^{-1}$ (NED) & 1910 & 1523      & 1821          & 2390      & 890       & 1749      & 1692 \\ 
Distance, Mpc (NED)  
        & 28.7         & 25.3      & 29.5		 & 30.8 	  & 7.3		  & 26.2 	   & 25.4\\ 
Inclination, 
deg (LEDA)  & 68.4       & 70.6      & 44.7          & 90.0      & 76.4      & 90.0      & 23.2  \\
{\it PA}$_{phot}$, deg 
(LEDA)     & 110       & 135     & 7.5           & 146       & 1.7       & 111.5     & -- \\ 
$\sigma _*$, $\mbox{km} \cdot\mbox{s}^{-1}$, (LEDA) 
		& --		 & -- 		 & 158 		 & 162	  & 50 	  & 167 	  & 112\\ 
$M_{HI}$, $10^9\,M_{\odot}$$^5$
		& --	 		 & $<$0.017 & --			 & 2.4 		  & -- 		  & $<$0.018  & 1.86\\
$M_{H_2}$, $10^8\,M_{\odot}$$^6$ 
		& --		 & $<$0.36 	 & -- 			 & 0.68 	  & --        & $<$0.42   & 5.89\\ 
\noalign{\smallskip}
Obs. date  &  12 Dec 2012 & 22 Apr 2012 & 20 Sep 2012 & 20 Nov 2011 &19 Nov 2011 & 12 Dec 2012 & 21 Apr 2012\\
Exp. time, sec  & 6000   & 2700      & 6600         & 5400       & 3600      & 8400      & 4500 \\
Seeing, arcsec  & 1.6    & 3.0       & 1.3          & 2.5        & 2.0       & 2.0       & 2.0 \\
\hline
\multicolumn{2}{l}{$^1$\rule{0pt}{11pt}\footnotesize NASA/IPAC Extragalactic
Database}\\ 
\multicolumn{2}{l}{$^2$\rule{0pt}{11pt}\footnotesize Third
Reference Catalogue of Bright Galaxies}\\
\multicolumn{2}{l}{$^3$\rule{0pt}{11pt}\footnotesize Lyon-Meudon Extragalactic
Database}\\ 
\multicolumn{2}{l}{$^4$\rule{0pt}{11pt}\footnotesize Makarov et al. data base based on LEDA and NED}\\ 
\multicolumn{2}{l}{$^5$\rule{0pt}{11pt}\footnotesize \citet{Serra2012_atlas3d13_HI}}\\ 
\multicolumn{2}{l}{$^6$\rule{0pt}{11pt}\footnotesize \citet{Young2013_atlas3d4_H2}}\\ 
\label{table_info_obs}
\end{tabular} 
\end{minipage}
\end{table*}

\section{Observations and data reduction}

Spectral observations of our sample of lenticular galaxies were carried out with
the multimode focal reducer SCORPIO-2 \citep{scorpio2} at the 6-m telescope of
the Special Astrophysical Observatory of the Russian Academy of Sciences.
Long-slit spectra of all galaxies besides NGC~6654 were acquired using
VPHG1200@540 grating which provided an intermediate spectral resolution
FWHM$\approx$4{\AA} in a wavelength region from 3800 to 7300\AA. The spectrum of
NGC~6654 was collected using the other grating, VPHG2300, which provided a bit
higher spectral resolution FWHM$\approx$2.2{\AA} in a shorter wavelength region
from 4800 to 5600\AA. All observations were taken with a $1\arcsec$-width slit
aligned along major axes of the galaxies. Observations of the galaxies with the
VPHG1200@540 grating were exposed using CCD 2k$\times$4k chip E2V CCD42-90 while
spectra obtained with the VPHG2300 were collected with EEV 42-40 2k$\times$2k
CCD. In both cases a scale along the slit was 0.357 arcsec pixel$^{-1}$. General
characteristics of the galaxies as well as observation dates, total exposure
times, and atmospheric seeing conditions are listed in Table~1.   

The primary  data reduction comprised the following steps: bias subtraction,
flat-fielding, removing cosmic ray hits using the Laplacian filtering technique
\citep{vanDokkum2001}, and building the wavelength solution using the He-Ne-Ar
arc-line spectra with accuracy of 0.1 and 0.05{\AA} for two grating setups. To
subtract the sky background, we  invented recently a rather sophisticated
approach \citep{KatkovChil2011_skysub}. We constructed model of the spectral
instrumental response of the spectrograph (LSF -- line-spread function) varied
along and across the wavelength direction by using the twilight spectrum. The
final stages of the long-slit spectra reduction were night sky spectrum
subtraction taking into account the LSF variations, spectra linearization, and,
to take into account spectral sensitivity variations, flux calibration by using
the spectrum of a spectrophotometric standard star.  The uncertainty frames were
initially computed from the photon statistics and the read-out noise values, and
then processed through exactly the same data reduction steps in order to
estimate the flux accuracy.

\subsection{Stellar kinematics}

\begin{figure*}
\centering
\includegraphics[width=0.33\textwidth]{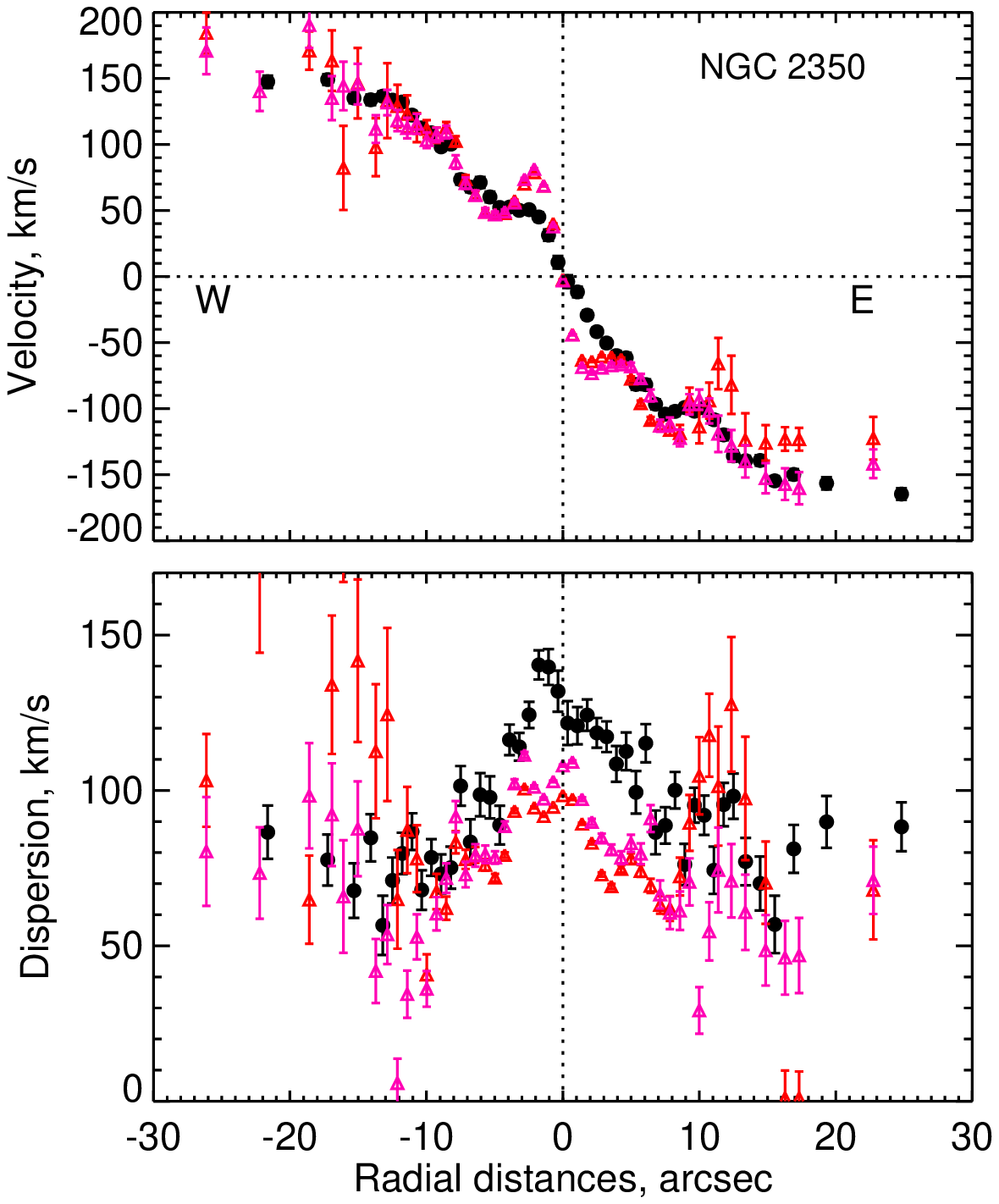}
\includegraphics[width=0.33\textwidth]{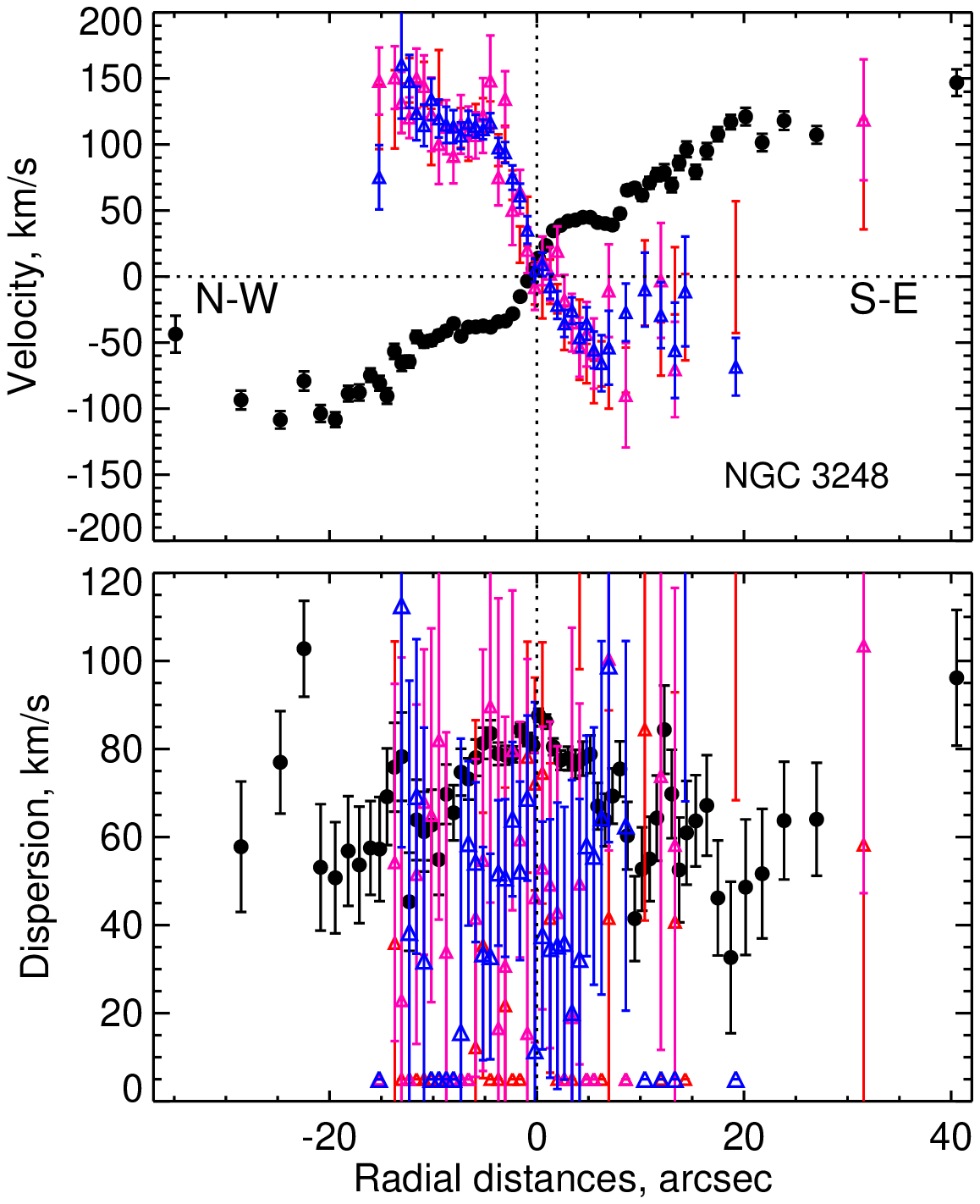}
\includegraphics[width=0.33\textwidth]{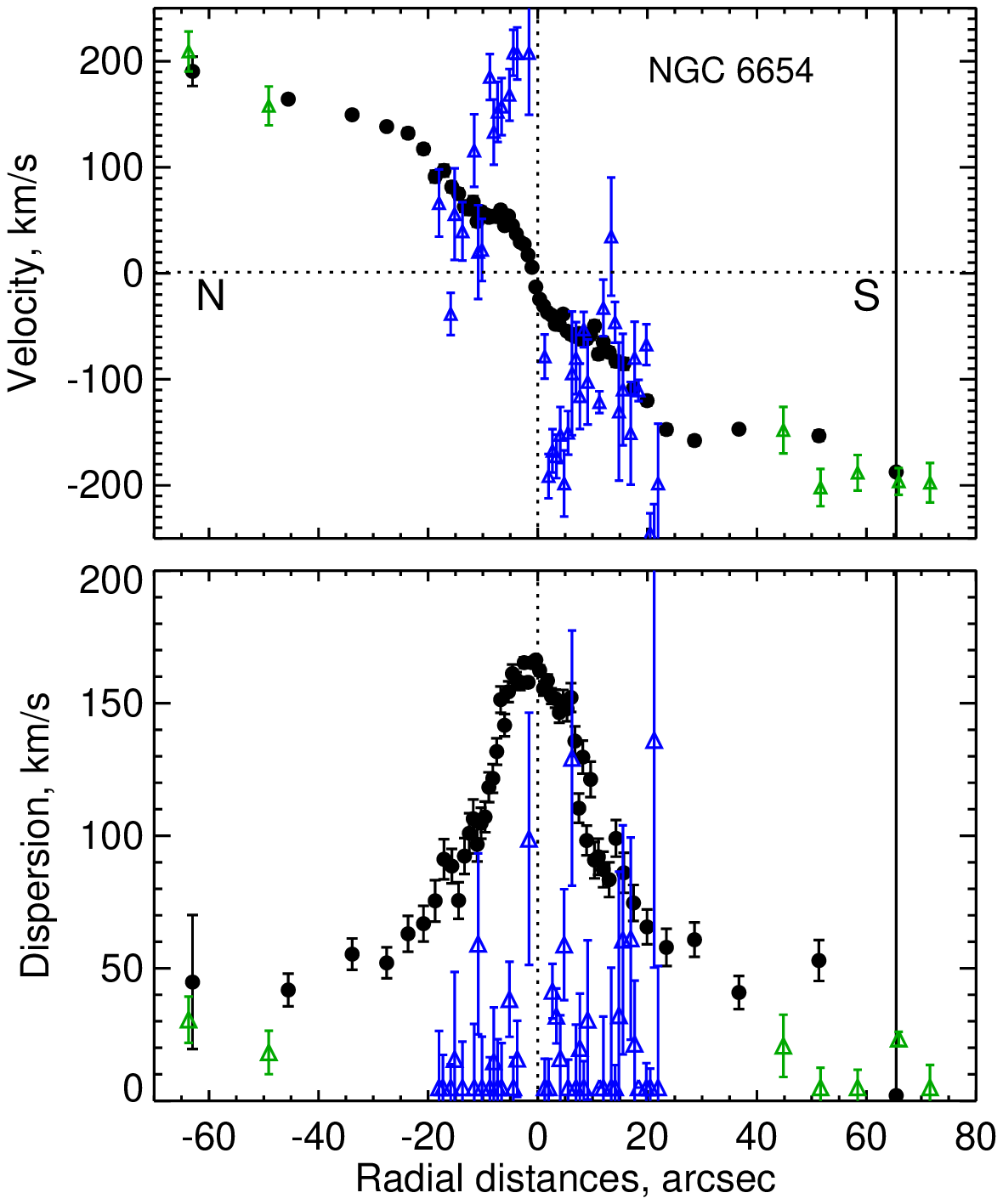}\\
\includegraphics[width=0.33\textwidth]{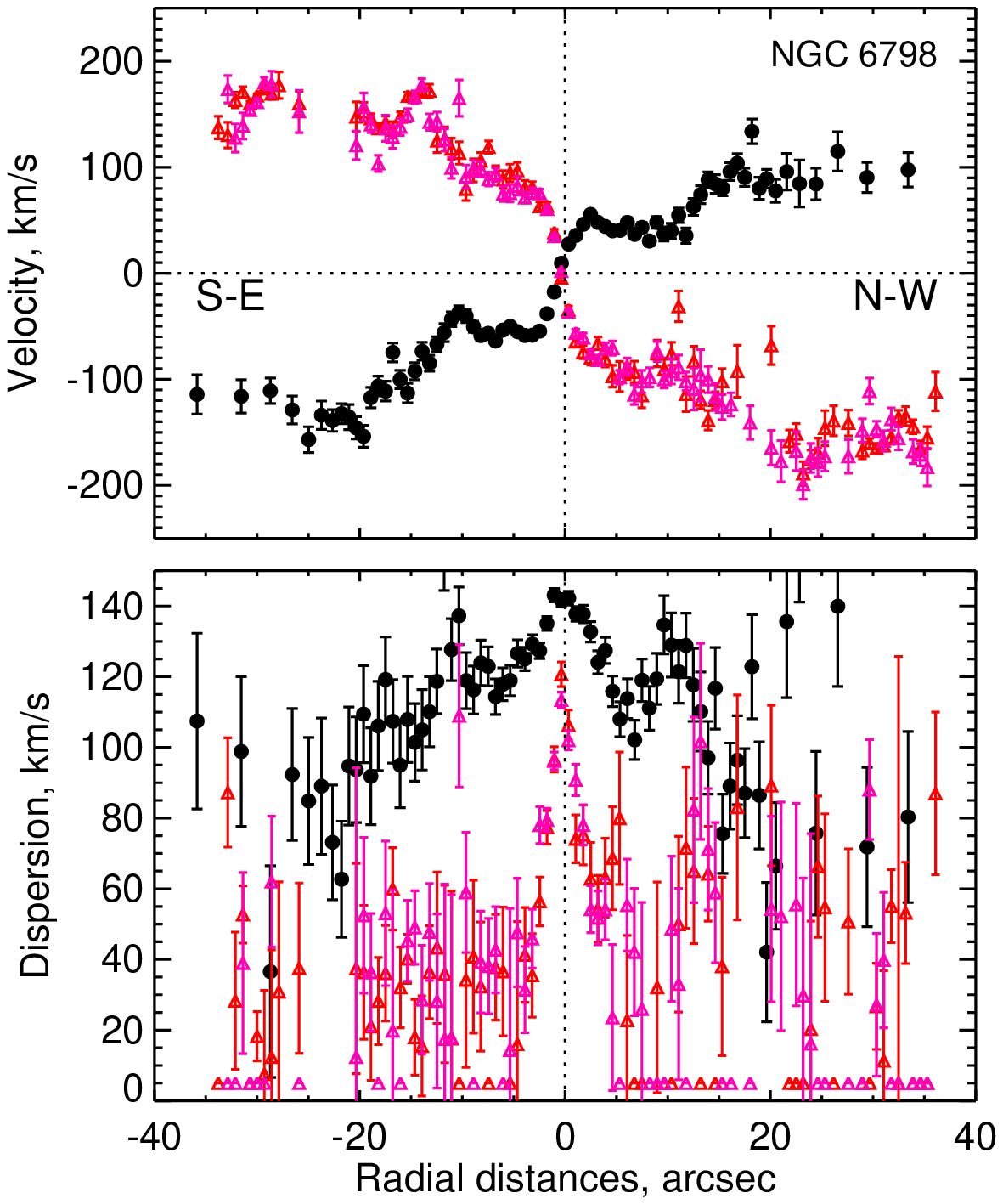}
\includegraphics[width=0.33\textwidth]{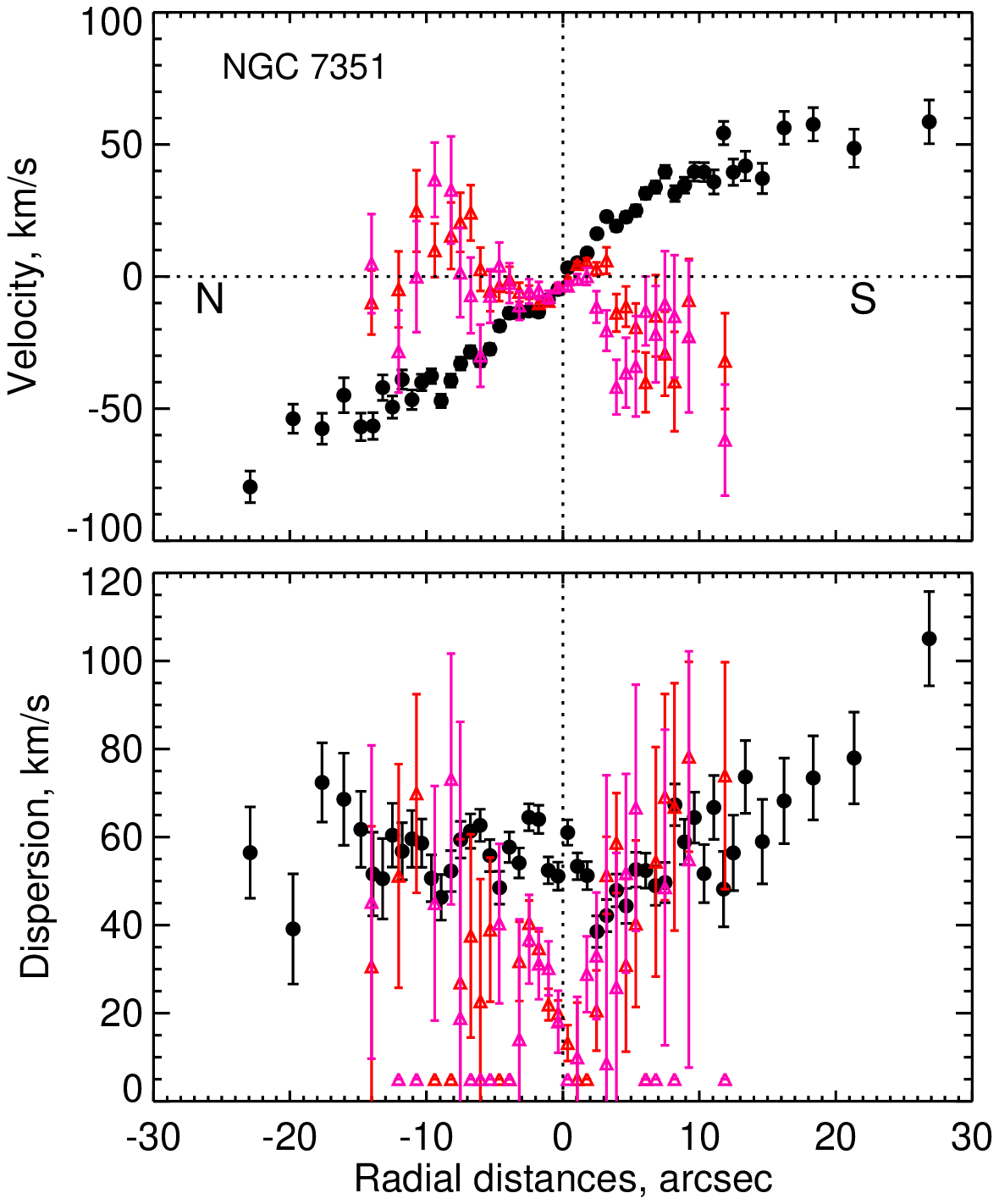}
\includegraphics[width=0.33\textwidth]{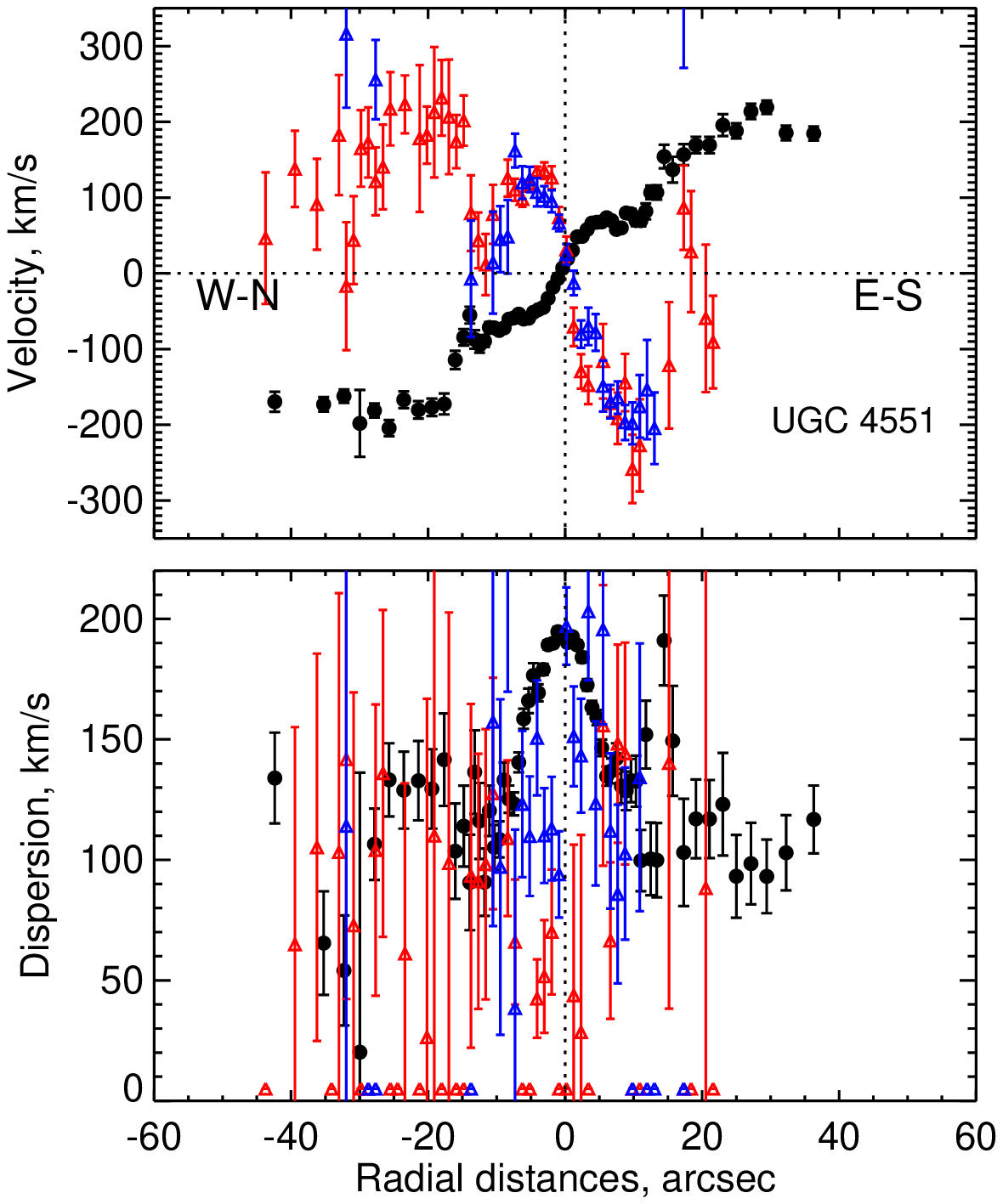}\\
\includegraphics[width=0.33\textwidth]{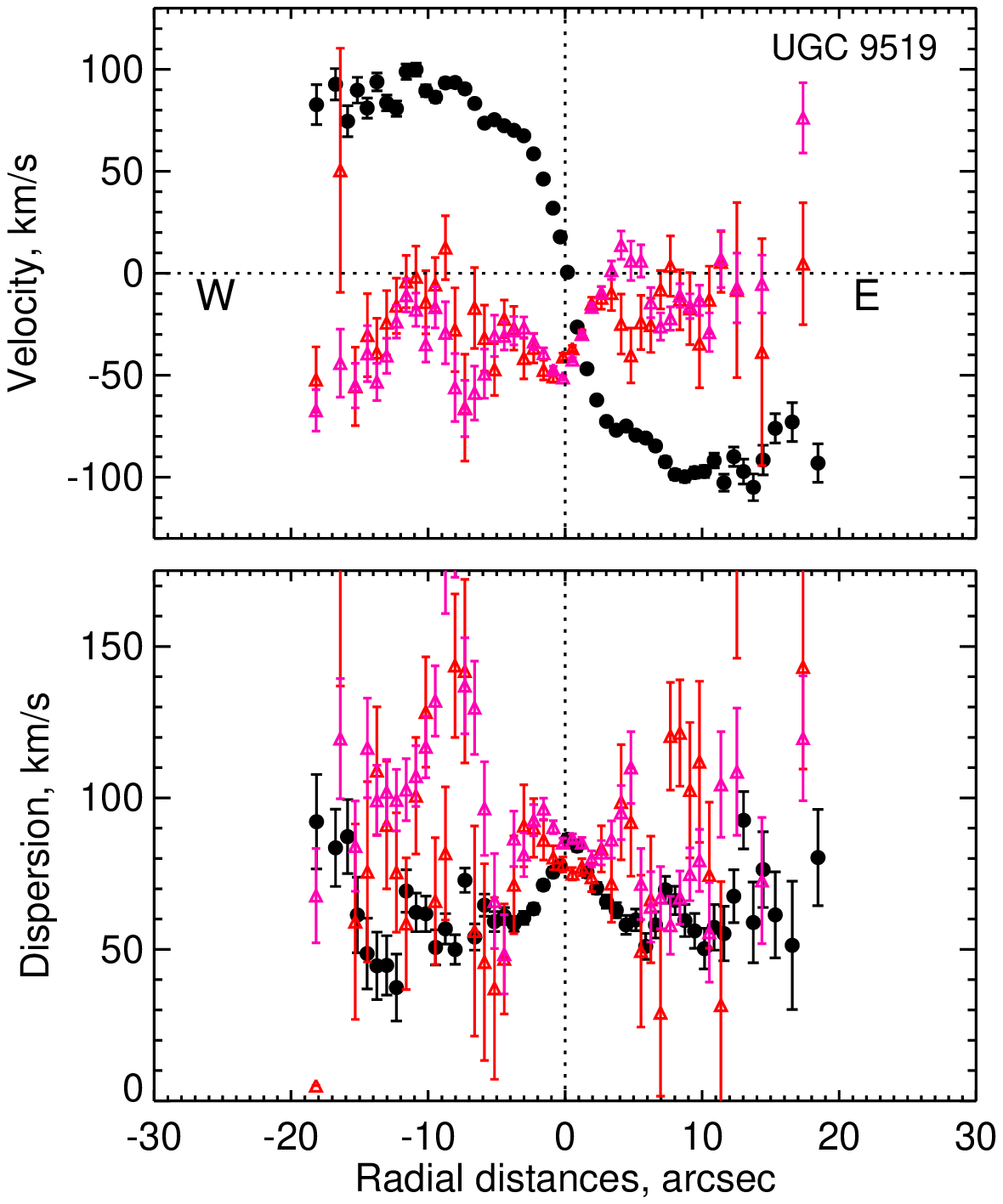}
\caption{Kinematical profiles of stars (black symbols) and ionized gas (colour
symbols) along the major axes. Different colours of symbols correspond to
kinematics measured in different emission lines: \Ha - red, [N\ii] - magenta,
[O\iii] - blue, and \Hb - green. The low velocity dispersions falling below
the possibility of being measured with our spectral resolution are indicated
as 5\km velocity dispersions.}
\label{pics_kinematics}
\end{figure*}

To derive information about stellar and ionized gas kinematics we first fitted
the stellar absorption spectra by the PEGASE.HR high resolution stellar
population models \citep{LeBorgne2004} convolved with a parametric line-of-sight (LOS)
velocity distribution by applying \textsc{nbursts} full spectral fitting
technique \citep{Chilingarian2007A, Chilingarian2007B}. Before the minimization
procedure, the model grid of stellar population spectra is convolved with the
LSF. Multiplicative Legendre polynomials are also included to take into account
possible internal dust reddening and residual spectrum slope variations due to
the errors in the assumed instrument spectral response.  Ionized-gas emission
lines and remnants of the subtracted strong airglow lines do not affect the
solution due to masking the narrow 15{\AA}-wide regions around them. The
resulted stellar parameters are LOS velocity $v$, velocity dispersion
$\sigma$, higher-order Gauss-Hermite moments $h_3$, $h_4$, and the stellar
population parameters: the age $T$ and metallicity [Z/H]. In this paper we
discuss only kinematics, while the stellar population properties will be
considered in the forthcoming publication. LOS velocities and
velocity dispersions for the stars derived along the major axes are shown in
Fig.~\ref{pics_kinematics}. 

\subsection{Ionized gas} 

The forthcoming analysis concerns optical emission lines of the ionized gas. We
subtracted the best-fit stellar model from the observed spectrum and fitted the
remaining pure emission lines by Gaussians pre-convolved with the spectrograph
LSF. In such a way, we have measured LOS gas
velocities and emission-line fluxes varying along the slit. Radial profiles of
gaseous kinematics are also shown in Fig.~\ref{pics_kinematics}. In all galaxies
where emission lines were detected, the derived profiles are rather extended
both for the stars and ionized gas. 

\begin{figure*}
\centering
\includegraphics[width=0.3\textwidth]{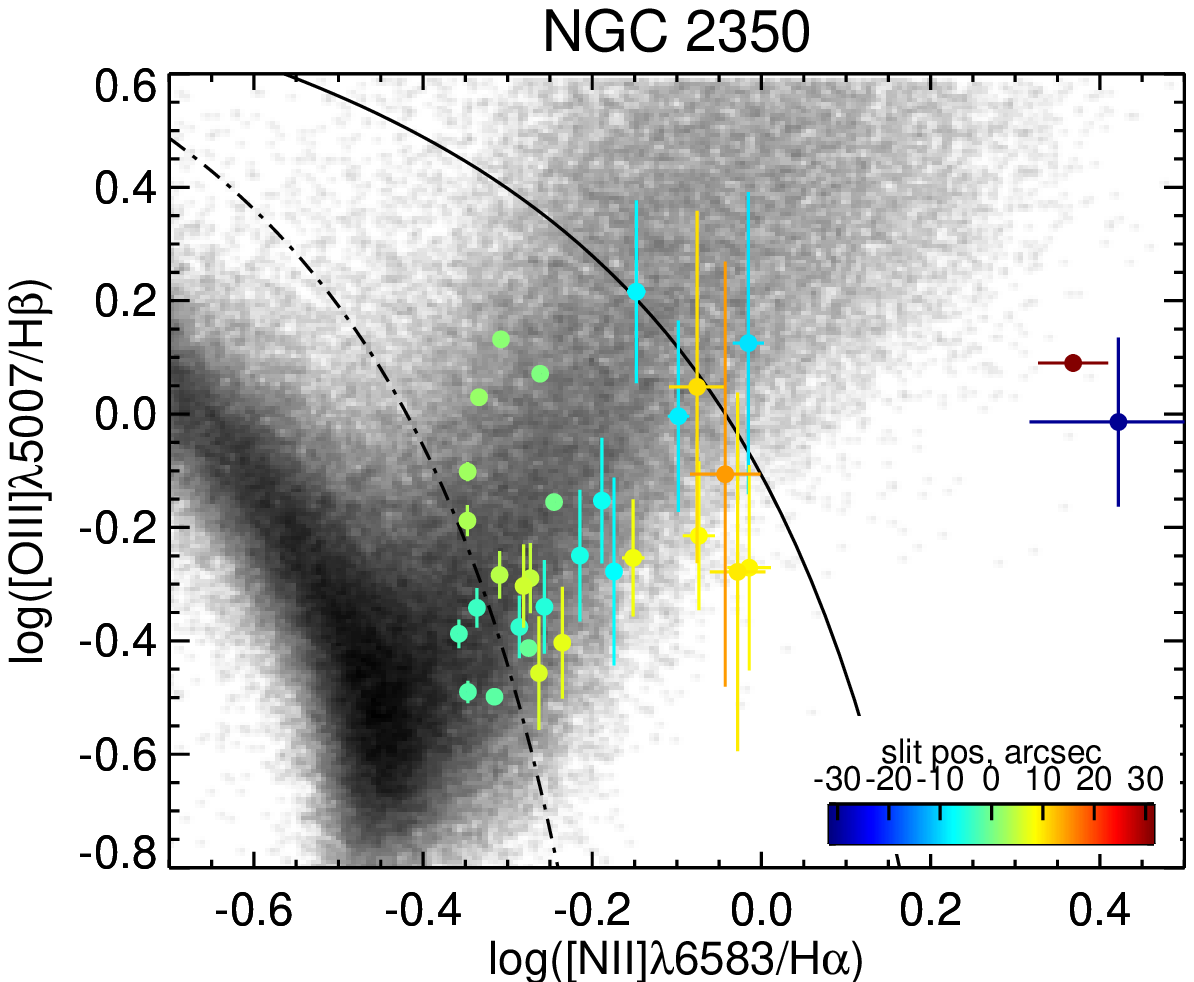}
\includegraphics[width=0.3\textwidth]{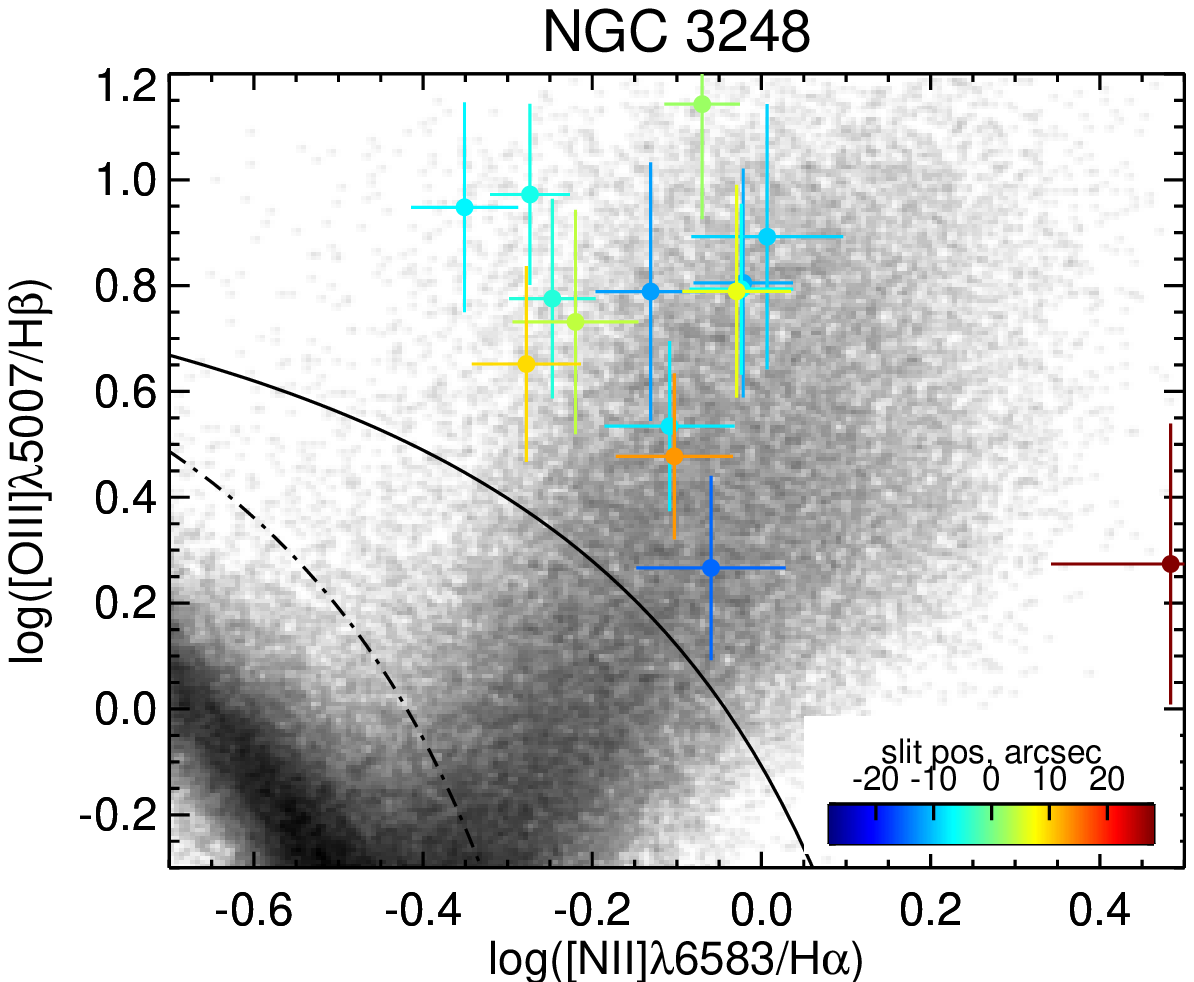}
\includegraphics[width=0.3\textwidth]{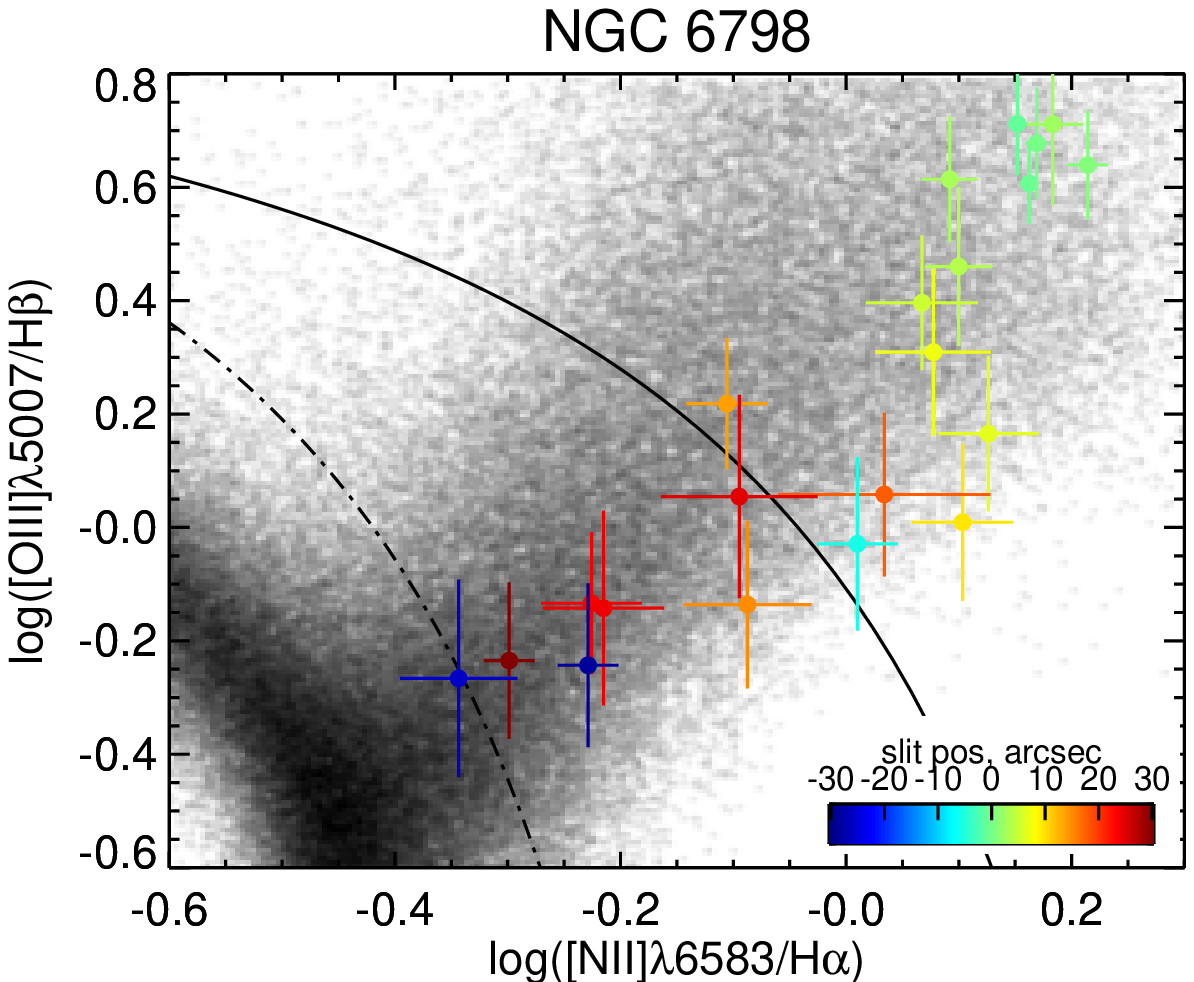}\\
\includegraphics[width=0.3\textwidth]{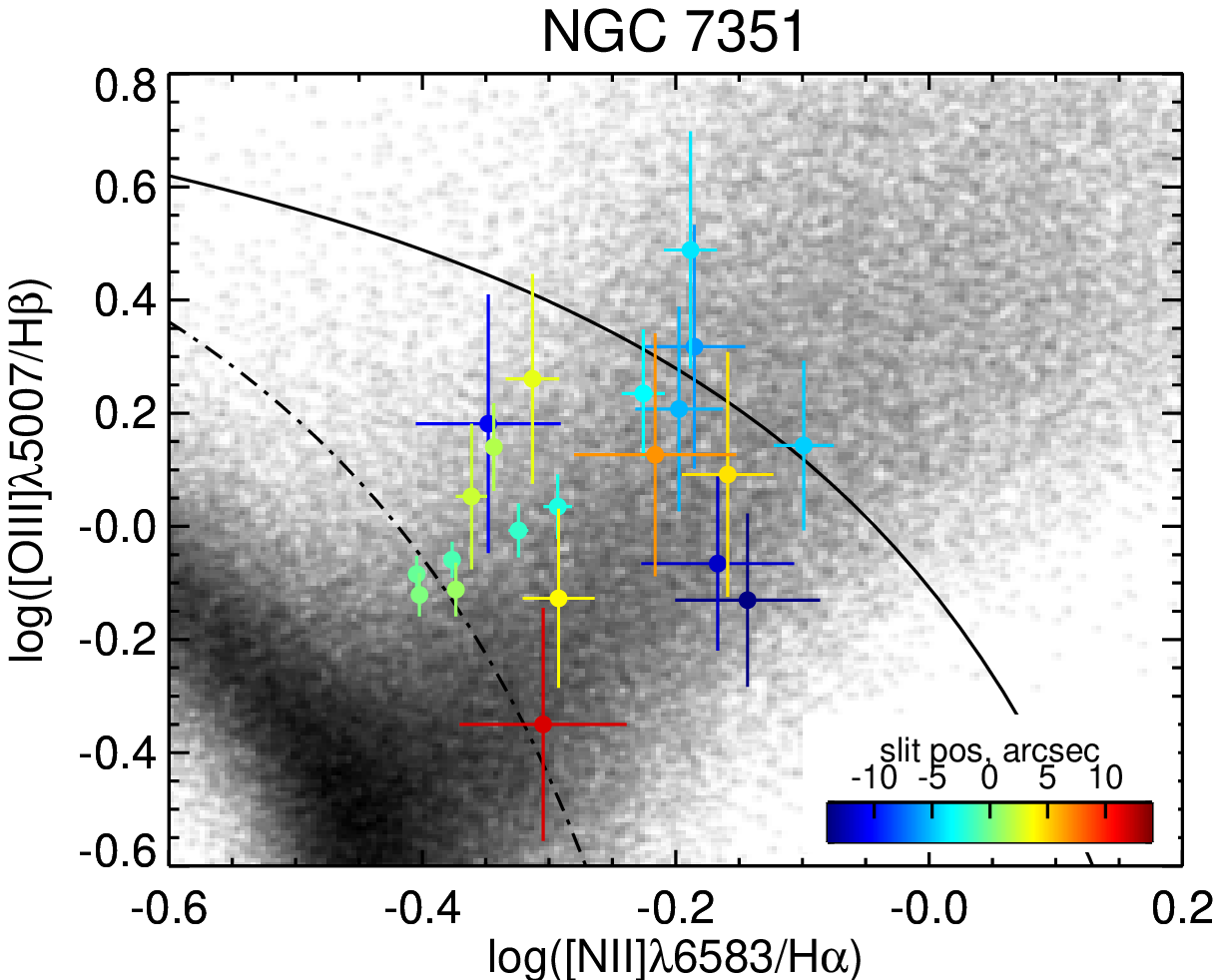}
\includegraphics[width=0.3\textwidth]{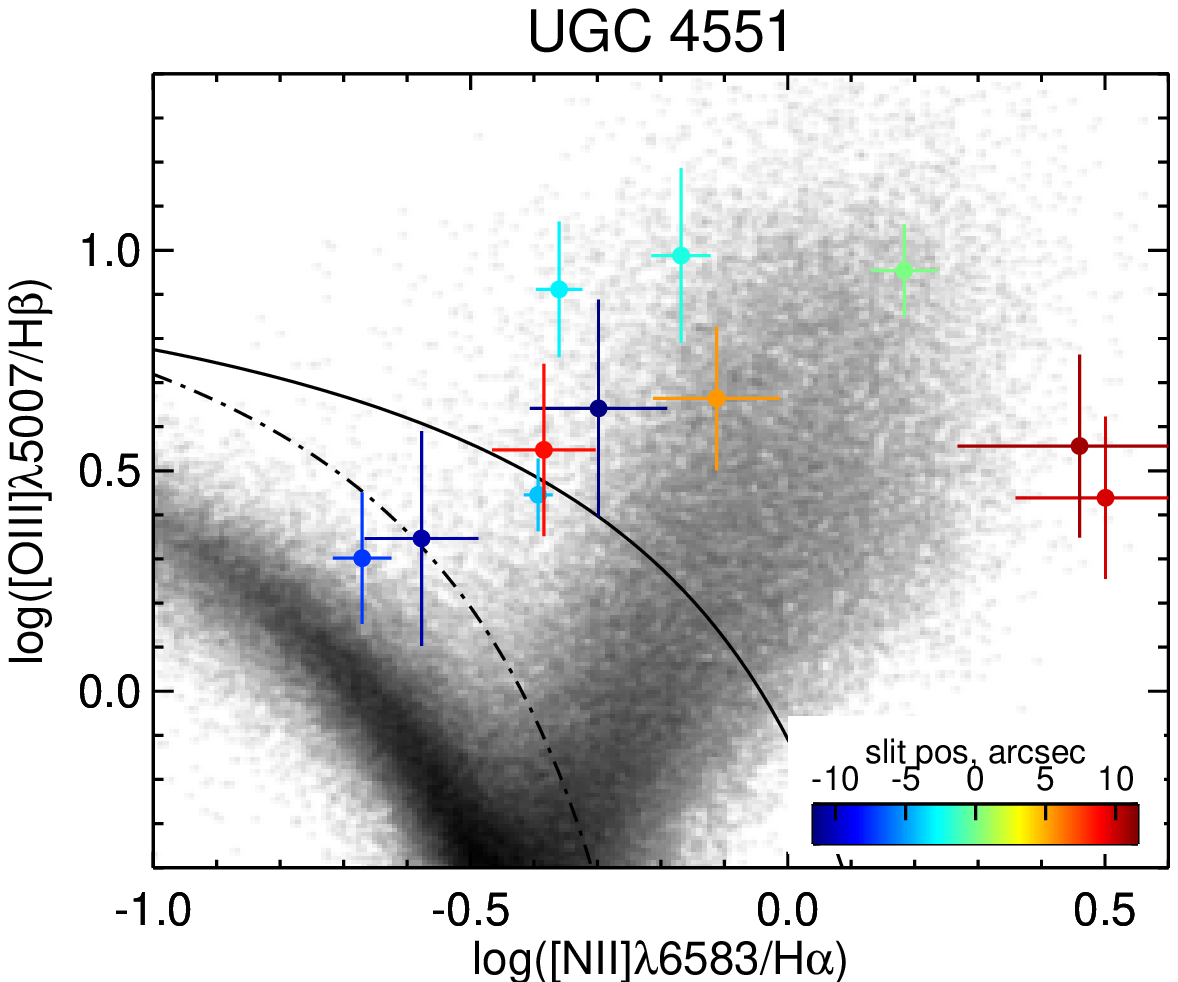}
\includegraphics[width=0.3\textwidth]{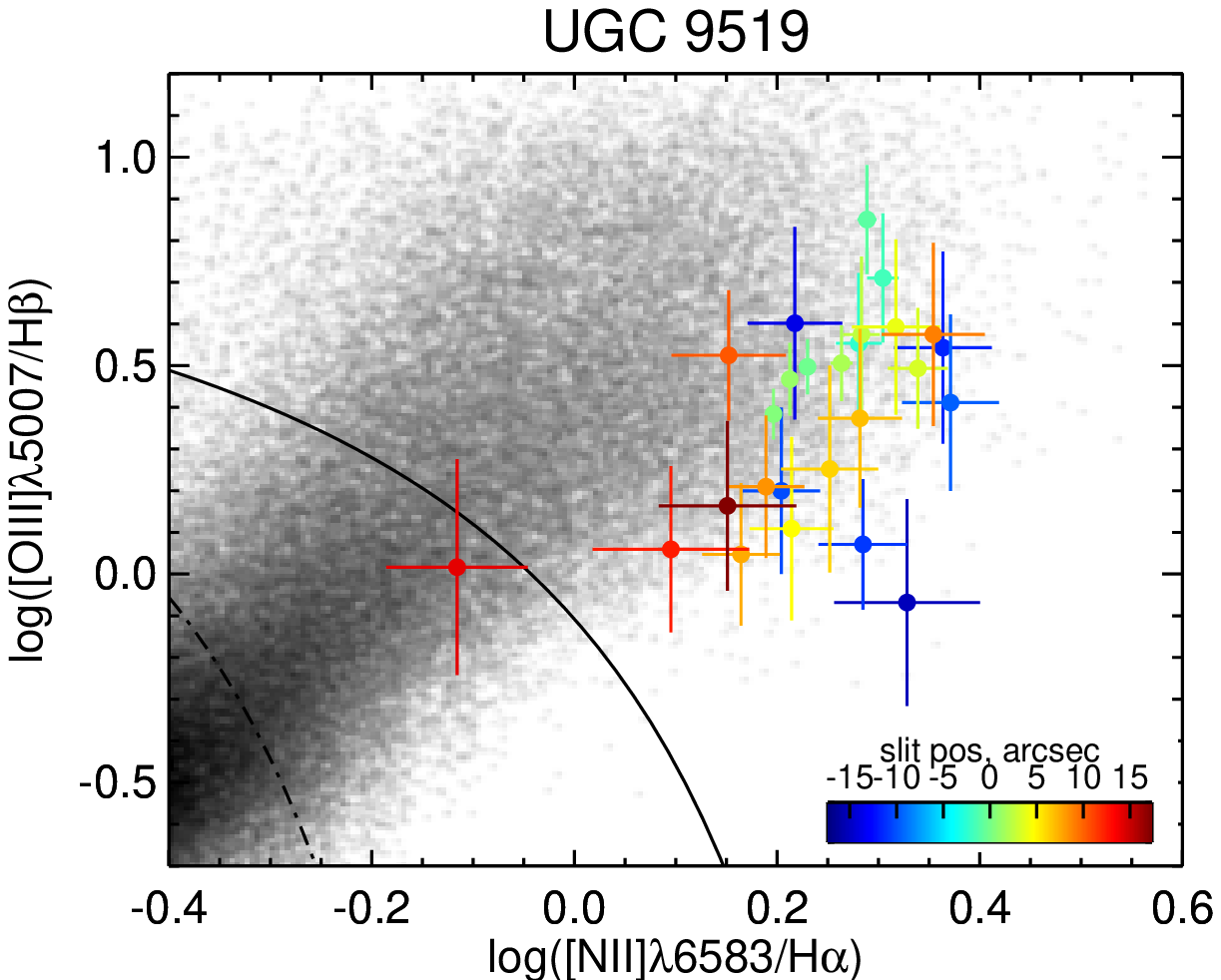}
\caption{Excitation diagnostic diagrams comparing the emission-line ratios:
[N\ii]/\Ha vs. [O\iii]/\Hb. The colour of the points codes the distance
from the galaxy centres.  The distribution of the measurements of the line
ratios for galaxies from the SDSS survey with high signal-to-noise ratios
(S/N$>$3 in every line) are shown by grey colour. The black curves, which
separate the areas with the AGN/LINER excitations (to the right) from areas 
with the SF-induced excitation (to the left from these curves), 
are taken from \citet{Kauffmann2003}
(dash-dotted curve) and from \citet{Kewley2006} (solid curve).}
\label{diagn_diags}
\end{figure*}

To identify the dominant source of the gas ionization, we plot our measurements
of the emission-line fluxes on to the classical diagnostic BPT-diagram
\citep{Baldwin1981bpt} (Fig.~\ref{diagn_diags}). In order to derive the line
ratio estimates for the BPT-diagram we needed higher signal-to-noise ratios than
ones required only to constrain kinematics. As a result, we have plotted
measurements where the emission lines \Hb, [O\iii], \Ha and [N\ii] are detected
all with S/N$>$3. Symbol colours code the distance from the centres of the
galaxies.

Almost all measurements in NGC~2350 and NGC~7351 as well as in the outer ring of
NGC~6798 are located in the BPT-diagram region where star-formation excitation
mechanism dominates. Emission-line measurements for the other galaxies --
NGC~3248, UGC~9519, and the most part of UGC~4551 -- fall into the
AGN/LINER-dominated (or shock-dominated) region of the BPT-diagram. 

\section{Results: Counterrotation}

We found that in our small sample of twelve galaxies observed with the SCORPIO-2
seven of them (7/12; 58\% $\pm$ 14\%) have revealed extended emission lines; and
among those, five galaxies (5/7; 71\% $\pm$ 17\%) demonstrate decoupled gas
kinematics with respect to their stellar components over the whole discs. Since
we have only long-slit measurements, we characterize the decoupled gas
kinematics as `counterrotation', though in some cases very different amplitudes
of the LOS velocity variations for the cold stellar component with negligible
asymmetric drift and ionized gas along the major axes (lines of nodes of the
stellar discs) indicate different planes of gas and star rotation. 

{\bf NGC~2350} is characterized by gas emission extending up to 25 arcsec
($\approx$ 3.5 kpc); and all the gas co-rotates the main
stellar disc. The gas has two peaks at velocity dispersion profile at radii
$r=-17$ arcsec and $r=12$ arcsec, where the galaxy image reveals bar ``ansae''
features.

In {\bf NGC~3248} the gas LOS velocity profile has asymmetric shape: the
North-West side has higher velocity amplitude than the South-East one. Probably
this asymmetry is associated with unsettled state of the gas subsystem.
This galaxy is included into the ATLAS-3D sample and is studied with the IFU
spectrograph SAURON. The maximum rotational velocity reached within the SAURON field
of view, $k_1=87.4$\km \citep{atlas3d_2_krajnovic2011}, is consistent with our long-slit
measurements. \citet{atlas3d_10} detected misalignment between the ionized gas
and stars ($\psi_{ion-star}=179.5\pm9.8$ degree) in NGC~3248. Unfortunately,
the gas velocity fields are unavailable for the most ATLAS-3D galaxies, and we cannot
confirm our findings of velocity asymmetries using the IFU data. However, the
similar positional angles of the kinematical major axes for the gaseous and stellar 
velocity fields, though indicating against radial inflows, do not exclude off-plane 
gaseous motions.

\begin{figure}
\centerline{
\includegraphics[width=0.25\textwidth]{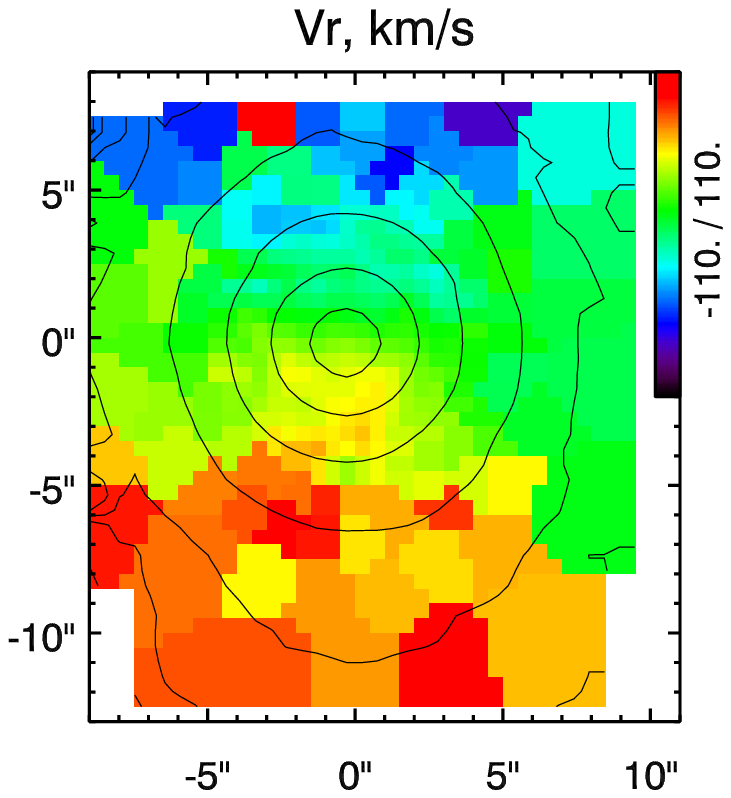}
\includegraphics[width=0.25\textwidth]{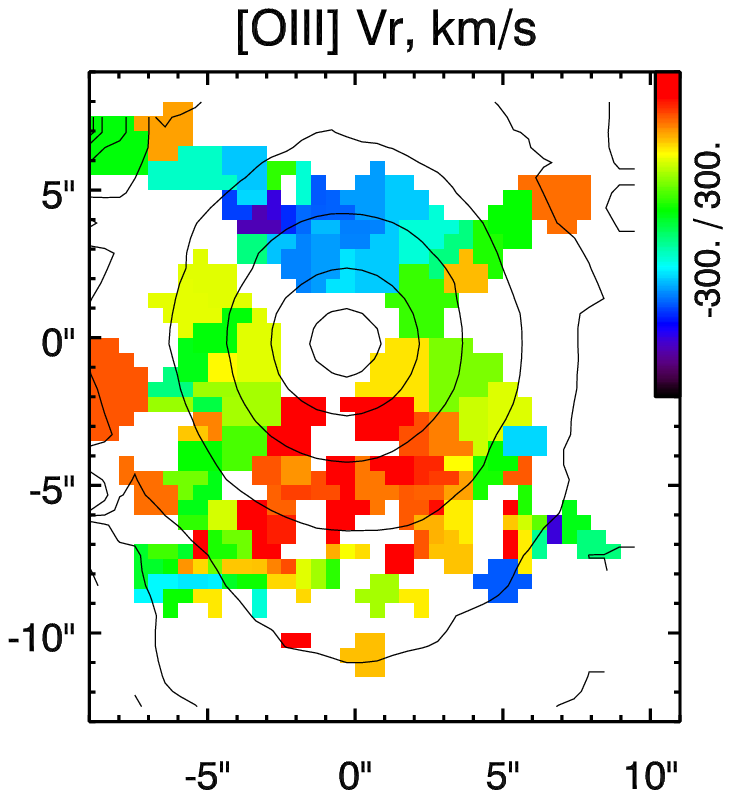}}
\caption{The results of analysis of MPFS data cube for NGC~6654. Stellar LOS velocity map  
is shown at \textit{left} panel, while the ionized gas velocity map measured in [O\iii] line 
-- at \textit{right} panel. Note that the amplitudes of velocities are very different.}
\label{MPFS}
\end{figure}

{\bf NGC~6654} is an example of weak emission lines. This galaxy was observed
under instrumental setup covering only \Hb and [O\iii] lines. [O\iii] doublet
shows intense increase of the relative LOS velocity amplitude up to 200--250\km in the
central few arcsec being perhaps decoupled though not counterrotating with
respect to stars, and then falls to the stellar rotation velocity level at a radial
distance of 10--15 arcsec. Kinematics of the \Hb emission line follows the stellar
rotation; this line is visible only in a few spatial bins at large radial
distances, $R>40$ arcsec. NGC~6654 was studied as a double-barred
candidate by \citet{Moiseev6654} who used the Multi-Pupil Fiber Spectrograph
(MPFS) at the 6-m Russian telescope. Moiseev has shown that there is no non-circular
motions in the stellar velocity field on the scale of a photometric secondary bar
which was then expected in N-body simulations. We have reanalyzed the MPFS science-ready 
cube for NGC~6654, which has been kindly provided by Alexey Moiseev, in the same
manner as our long-slit data. The stellar velocity field and velocities of the [O\iii]
emission line are presented in Fig.~\ref{MPFS}. One can see that ionized gas
has a similar kinematical major axis as the stars but the velocity amplitudes are quite
different and both are consistent with our long-slit measurements. To explain extreme
visible velocity amplitudes of the ionized gas, we have considered a possibility
of planar non-circular gas motions within triaxial potential of a nuclear bar with
a radius of 4 arcsec which is known in this galaxy \citep{erwinsparke}. However, 
the orientations of both bars in NGC~6654, $PA=13^{\circ}$ for the primary one and 
$PA=135^{\circ}$ for the secondary one \citep{nirs0s}, differ substantially from 
the disc line of nodes orientation and are not orthogonal to it, so the gas affected 
by these bars must demonstrate the turn of the kinematical major axis away from the 
disc line of nodes. Since we do not observe such a turn, we conclude that the only possible
explanation of the fast visible gas rotation is that we look at the gaseous disc under
larger inclination than at the stellar disc of the galaxy; so the stellar and gaseous
discs are again decoupled.

Another galaxy with a very extended gaseous disc is {\bf NGC~6798}. However in
this galaxy the gas counterrotates with respect to the stars; our measurements
reach the radius up to 35 arcsec (5.2 kpc) and reveal a slightly higher
amplitude of the gas rotation curve with respect to the stellar one.
Such behaviour is expected due to asymmetric drift taking into account
sufficiently high stellar velocity dispersion in the disc of this galaxy. The gas
counterrotation continues beyond the optical borders of the galaxy as HI
mapping shows \citep{Serra2012_atlas3d13_HI}. NGC~6798 also was
considered by \citet{atlas3d_10} where the authors calculated an angle between
the kinematical axes of stars and gas $\psi_{ion-star}=174.0\pm6.9$ degree.

{\bf NGC~7351}, a dwarf S0 galaxy, has also weak emission lines with relatively
large scatter of measurements; despite that the gaseous kinematical profile in
Fig.~\ref{pics_kinematics} clearly indicates the decoupled kinematical
properties of the ionized gas in this galaxy beyond the very central part,
$R>5\arcsec$. 

Ionized gas kinematics in {\bf UGC~4551} has a complex character. Relative LOS
velocities of the gas rise rapidly in the central region, then the amplitude falls
down to null and then the North-Western side of the velocity profile rises up to 200\km
with following declining.

{\bf UGC~9519} has emission-line structures extending (projecting?) all over
the stellar disc of the galaxy.  The low velocity amplitude of the gas when
compared to the stars indicates that the gas rotates in another plane than the
stars. The warped polar disc in this system has already been mapped by
\citet{Serra2012_atlas3d13_HI} in the 21 cm line of the H\i, the polar rotation 
of the molecular and ionized gas rotation is reported by
\citet{Alatalo_atlas3d_CO_2013} and \citet{atlas3d_10, Davis2013_atlas3d} .
The conclusion about the polar orientation of the gaseous disc in the very centre
of the galaxy is also supported by a circumnuclear red dust lane visible at
the SDSS images of the galaxy which is roughly perpendicular to the major axis
of the central isophotes (to the bar?); the whole picture is consistent with
the gaseous disc, polar in the centre and strongly warped farther from the
centre. Let us note that systematic velocities of the gaseous and stellar
subsystems differ by 30--40\km, and the symmetry centre of the gas velocity
profile is also shifted with respect to the galactic centre along the slit.
Perhaps, the gas subsystem of UGC~9519 was accreted recently and has not
settled down yet.

\section{Discussion and Conclusions}

We have studied a sample of 12 highly isolated S0 galaxies by means of
long-slit spectroscopy undertaken with the reducer SCORPIO-2 of the Russian 6m
telescope.  In seven galaxies we have found extended gaseous discs, and in 5
galaxies the extended emission-line gas counterrotates their stellar
components. The gas excitation is found to be mostly shock-like though a few
starforming rings may be suspected.

By comparing our results on the frequency of extended ionized-gas discs in S0
galaxies with the earlier statistics, we see full agreement:
\citet{kuijken_fisher} found ionized gas in 17 of 29 S0s studied, so their
fraction of gas-rich S0s is about 58$\pm$9\%, just as in our study. However, if
we consider a fraction of counterrotating gaseous discs among all extended
gaseous discs in S0 galaxies, we see a prominent difference. When S0 galaxies
were selected over all types of environment, the fraction of counterrotating
gaseous discs was 20\%--24\%\ \citep{bertola92,kuijken_fisher}; more exactly,
by combining two samples, \citet{kuijken_fisher} gave 24\%$\pm 8$\%.  In our
sample, the fraction of counterrotating gaseous discs is 71\%$\pm 17$\%.  We
expected such a trend because \citet{atlas3d_10} noted a dependence of gas
kinematics in the early-type galaxies (mostly S0s in the ATLAS-3D sample) on 
their environment: dense environment provided
tight coincidence between gas and star kinematics while in more sparse
environments the fraction of decoupled gaseous kinematics grew. Our isolated S0
galaxies represent an extreme point in this dependency, and the fraction of
decoupled gas kinematics exceeds 50\%. Following the logic of
\citet{bertola92}, according to which when gas is accreted, it should have an
arbitrary spin, and so long-slit spectroscopy should reveal an equal fraction
of corotating and counterrotating gaseous discs. If one takes into
account the possibility for internal gas incidence, the fraction of systems with
counterrotating gaseous discs would be less than 50\%. But our results based on
the sample of galaxies in the strictly sparse environment suggest that the fraction
of the gas counterrotations is higher than 50\%. Hence, we can conclude that
in isolated S0s their gas is virtually always accreted from external sources!
But to be certain that the directions of possible gas accretion are distributed
isotropically, we must firstly identify sources of gas accretion. Our galaxies
are isolated so they cannot acquire their gas from neighbours of comparable
mass/luminosity; the sources of cold gas accretion may be dwarf satellite
merging \citep{Kaviraj2009,Kaviraj2011} or perhaps cosmological gas filaments \citep{keres_flows,
dekel_flows}. Are they distributed isotropically? This is a good question.

\subsection{Dichotomy of gas excitation.}

It is clear, though not from long-slit data along major axes alone, that
generally the gaseous discs in S0s are not coplanar to the stellar ones; so
what we call `counterrotation' may be projection of an inclined gaseous disc
onto the stellar disc plane. Perhaps there exists a dichotomy concerning the
mechanisms of ionized-gas excitation in the discs of S0 galaxies. 

\citet{wakamatsu93} showed that the shocks can be generated when gas on
polar orbits crosses the potential well of a stellar disc, like the
grand-design shock waves in spirals or along the bars of barred galaxies
are generated by stellar density enhancements. In this sense the inclined
gaseous discs are similar to polar ring/disc structures; so in the cases of 
gas motions in inclined planes the shock-like excitation of the gas is expected, and
emission-line ratio measurements should be found in the LINER region of the BPT
diagram. Another mechanism providing LINER-like emission is ionization by
evolved stars during very hot and energetic post-AGB phase. It is thought to be
especially important in early-type galaxies \citep{Sarzi2010,
Singh2013_califa_postAGB, Bremer2013}. Both mechanisms ionize the gas without
involving the radiation field of young massive stars. It is expected
that both mechanisms would lead to ionization only at the region where gas
crosses the galaxy plane due to short cooling time compared to dynamical time.
The gradient in excitation should be visible in IFU data.

When the gas is accreted smoothly in the plane of a stellar disc, there are
more possibilities to conserve its coolness with following ignition of star
formation. 
The location of emission-line ratio measurements in the SF region of the BPT
diagram supposes that the young stars radiation contributes the dominant source
of gas ionization.  

Through the observational data presented here we see that in the galaxies where
the gas is probably confined to the disc planes (NGC~2350, NGC~6798, NGC~7351,
the very outer part of NGC~6654) the excitation by young stars is preferable.
Shocks or post-AGB stars are the main agent of gas excitation in
the remaining galaxies (NGC 3248, UGC 4551, UGC 9519) whose velocity profiles
indicate asymmetries and complex features which are probably resulted from the
gas motions in the inclined planes. 

Earlier we observed this dichotomy when we
found counterrotating gas in S0 galaxies NGC~2551 and NGC~5631
\citep{n2551_n5631}: the coplanar discs of NGC~2551 look UV-bright in the GALEX
data so intense star formation proceeds over the counterrotating gaseous disc
in this galaxy, while in NGC~5631 the gaseous disc is inclined, and there is no
prominent signs of  star formation in it but the emission-line ratios all over the
disc demonstrate the excitation dominated by shocks or old post-AGB stars. Similarly,
the shock-like excitation of the emission-line gas was found by us in the
inclined gaseous disc of S0 galaxy NGC~7743 \citep{n7743}. One more example of
current star formation in the counterrotating gaseous disc coplanar to a main
stellar disc represents the lenticular galaxy IC~719 reported by us recently
\citep{ic719}; earlier this galaxy was observed in the frame of the ATLAS-3D 
project by Alatalo et al.(2013).

\section*{Acknowledgements}

We thank Alexei Moiseev for providing a part of observations at the 6m telescope
and science ready MPFS cube of NGC~6654. The authors thank the anonymous referee
for constructive advices that helped us improve the paper.  The work was
supported by the Russian Federation President's grant MD-3288.2012.2 and Russian
Foundation for Basic Research (projects no.  13-02-00059a, 12-02-31452). IYK is
grateful to Dmitry Zimin's non-profit Dynasty Foundation.

\bibliographystyle{mn2e}
\bibliography{katkov}

\label{lastpage}

\end{document}